\documentclass[nonacm, natbib]{acmart}

\usepackage{tabularx,ragged2e,adjustbox,booktabs}
\renewcommand{\arraystretch}{1.06}
\newcolumntype{Y}{>{\RaggedRight\arraybackslash}X}

\AtBeginDocument{%
  }

\setcopyright{acmlicensed}
\copyrightyear{2026}
\acmYear{2026}
\acmDOI{XXXXXXX.XXXXXXX}
\acmConference [WSESE'26]{Make sure to enter the correct
  conference title from your rights confirmation email}{April 03--05,
  2026}{Rio de Janeiro, Brazil}
\acmISBN{978-1-4503-XXXX-X/2018/06}

\begin{document}

\title{LLM-Assisted Thematic Analysis: Opportunities, Limitations, and Recommendations}

\author{Tatiane Ornelas}
\affiliation{%
      \institution{Pontifical Catholic University of Rio de Janeiro}
      \city{Rio de Janeiro}
      \state{RJ}
      \country{Brazil}
  }
  \email{talves@inf.puc-rio.br}

\author{Allysson Allex Araújo}
\affiliation{%
  \institution{Federal University of Cariri}
  \city{Juazeiro do Norte}
  \state{CE}
  \country{Brazil}
}
\email{allysson.araujo@ufca.edu.br}

\author{Júlia Araújo}
\affiliation{%
      \institution{Pontifical Catholic University of Rio de Janeiro}
      \city{Rio de Janeiro}
      \state{RJ}
      \country{Brazil}
  }
  \email{jcaraujo@inf.puc-rio.br}

\author{Marina Araújo}
\affiliation{%
      \institution{Pontifical Catholic University of Rio de Janeiro}
      \city{Rio de Janeiro}
      \state{RJ}
      \country{Brazil}
  }
  \email{maraujo@inf.puc-rio.br}

\author{Bianca Trinkenreich}
\affiliation{%
      \institution{Colorado State University}
      \city{Fort Collins}
      \state{Colorado}
      \country{USA}
  }
  \email{bianca.trinkenreich@colostate.edu}

\author{Marcos Kalinowski}
\affiliation{%
      \institution{Pontifical Catholic University of Rio de Janeiro}
      \city{Rio de Janeiro}
      \state{RJ}
      \country{Brazil}
  }
  \email{kalinowski@inf.puc-rio.br}

\renewcommand{\shortauthors}{Ornelas et al.}

\begin{abstract}
[Context] Large Language Models (LLMs) are increasingly used to assist qualitative research in Software Engineering (SE), yet the methodological implications of this usage remain underexplored. Their integration into interpretive processes such as thematic analysis raises fundamental questions about rigor, transparency, and researcher agency. [Objective] This study investigates how experienced SE researchers conceptualize the opportunities, risks, and methodological implications of integrating LLMs into thematic analysis. [Method] A reflective workshop with 25 ISERN researchers guided participants through structured discussions of LLM-assisted open coding, theme generation, and theme reviewing, using color-coded canvases to document perceived opportunities, limitations, and recommendations. [Results] Participants recognized potential efficiency and scalability gains, but highlighted risks related to bias, contextual loss, reproducibility, and the rapid evolution of LLMs. They also emphasized the need for prompting literacy and continuous human oversight. [Conclusion] Findings portray LLMs as tools that can support, but not substitute, interpretive analysis. The study contributes to ongoing community reflections on how LLMs can responsibly enhance qualitative research in SE.

\end{abstract}




\maketitle

\section{Introduction}
Large Language Models (LLMs) are transforming how qualitative research is conducted in Software Engineering (SE) \cite{bano2024large}. Beyond their technical capabilities in code generation or documentation, they are also increasingly used to process, summarize, and interpret textual data from qualitative research \cite{BanoEtAl2023LargeLanguageModels, TrinkenreichEtAl2025GetTrainBe}. This shift has created new possibilities for scaling qualitative studies but, on the other hand, calls for reflection on how automation interacts with human interpretation, a cornerstone of qualitative research.

Among the diverse approaches to qualitative analysis, Thematic Analysis (TA) occupies a key role in SE research \cite{defranco2017content}. Rooted in interpretivist traditions, TA provides a systematic yet flexible method for identifying and interpreting patterns of meaning in qualitative data \cite{braun2021thematic}. Its iterative cycles of open coding, theme generation, and theme reviewing depend heavily on researchers’ theoretical sensitivity, reflexivity, and contextual understanding \cite{braun2019reflecting}. Consequently, introducing LLMs into these interpretive stages raises questions about how LLM assistance can coexist with the depth and authenticity that qualitative reasoning demands.

Recent studies suggest that LLMs can support key phases of qualitative analysis by accelerating coding, improving consistency, and enabling analysis of larger datasets \cite{BarrosEtAl2025LargeLanguageModel, TrinkenreichEtAl2025GetTrainBe, NaeemEtAl2025ThematicAnalysisArtificial}. When used carefully, LLMs can reduce repetitive work and promote transparency through traceable prompt-based procedures. However, these benefits come with challenges. LLMs are prone to hallucination, bias amplification, and contextual oversimplification \cite{SchroederEtAl2025LargeLanguageModels}. Hence, their outputs are highly sensitive to prompt design, requiring methodological competence in what is now termed prompting literacy. In other words, overreliance on automation can lead to premature closure of interpretation, weakening the iterative dialogue between researcher and data that characterizes qualitative analysis. Ensuring rigor and confidence in such contexts demands methodological safeguards, such as critical validation, documentation of prompts, and human oversight throughout the analytical process \cite{BaltesEtAl2025GuidelinesEmpiricalStudies}.

While initial studies in SE have discussed conceptual and methodological implications of using LLMs in SE research \cite{BaltesEtAl2025GuidelinesEmpiricalStudies, BarrosEtAl2025LargeLanguageModel, TrinkenreichEtAl2025GetTrainBe}, empirical investigations remain underexplored. Most implementations of LLM-assisted analysis to date have emerged in other fields, such as social sciences and healthcare, where researchers have explored how these models can support coding or theme development \cite{Paoli2024FurtherExplorationsLLM, NaeemEtAl2025ThematicAnalysisArtificial}. However, little is known about how experienced SE researchers themselves perceive this integration, including its potential opportunities, limitations, and methodological implications.

In this regard, the objective of this study is to examine how experienced researchers perceive the integration of Large LLMs into TA, focusing on potential opportunities, limitations, and methodological implications. For this reason, the International Software Engineering Research Network (ISERN) offers an ideal forum for this type of collective reflection since its members are actively engaged in advancing empirical methods and fostering methodological rigor across the SE community. To address this opportunity, we conducted a reflective workshop during the ISERN 2025 meeting, involving 25 researchers with substantial experience in empirical SE. Participants discussed how LLMs might support three core phases of the TA process (open coding, theme generation, and theme reviewing) using collaborative color-coded canvases to document reflections on opportunities, risks, and recommendations. To support the discussion, participants were introduced to a conceptualization of LLM-Assisted TA, describing how LLMs could be embedded into Braun and Clarke’s~\cite{BraunClarke2006UsingThematicAnalysis} TA phases as analytical partners through cyclical human–AI collaboration. This conceptualization provided a structured basis for discussion.


This study reports on a collective and highly qualified reflection about the use of LLMs in TA within SE and makes the following contributions: 1) An empirical synthesis of how experienced researchers perceive the opportunities and risks of using LLMs during the main phases of TA; 2) Observations on methodological tensions that arise when LLMs interacts with interpretive work, including effects on researcher roles and analytical depth; and 3) Considerations that may support future methodological reflection on the use of LLM-assisted TA.


\section{Background and Related Work}
\label{sec:background}

The integration of LLMs into qualitative research has become an emerging topic across disciplines, raising methodological and epistemological questions about rigor, transparency, and human interpretive control. This section reviews how studies have approached this integration from both methodological and practical perspectives. In SE, the discussion has evolved toward reflection and guidance, while in other domains, researchers have explored concrete applications of LLMs to the analytical stages of qualitative and TA. These directions provide the conceptual background for understanding the opportunities and challenges that motivate this study.


Within SE, early reflections emphasized that although LLMs expand the scale and efficiency of qualitative work, their use must remain grounded in methodological rigor and researcher reflexivity to preserve interpretive validity~\cite{BanoEtAl2023LargeLanguageModels}. Subsequent investigations documented efficiency gains and accessibility benefits but also concerns about interpretive depth, contextual sensitivity, and ethics~\cite{DeMoraisLecaEtAl2025ApplicationsImplicationsLarge, BarrosEtAl2025LargeLanguageModel}. More recent contributions consolidated potential evaluation guidelines and design principles that highlight transparency, reproducibility, and the continued need for human validation~\cite{WagnerEtAl2025EvaluationGuidelinesEmpirical, TrinkenreichEtAl2025GetTrainBe, BaltesEtAl2025GuidelinesEmpiricalStudies}. Overall, these works mark a transition from questioning whether LLMs belong in SE research to examining how they can be responsibly integrated through human-in-the-loop approaches.

Beyond SE, cross-disciplinary studies have also explored how LLMs can assist interpretive stages of analysis. Early experiments showed that models can perform coding and grouping with partial validity, but abstraction and theoretical grounding remain human-driven~\cite{Paoli2024PerformingInductiveThematic, Paoli2024FurtherExplorationsLLM}. Comparative analyses found that LLMs may generate a larger and more diverse set of codes while missing context or coherence, underscoring the need for reflexive interpretation~\cite{MathisEtAl2024InductiveThematicAnalysisHealthcare}. Collaborative systems such as \textit{CollabCoder} embed transparency and traceability~\cite{GaoEtAl2024CollabCoder}, and structured toolkits show how LLMs can support all TA phases when guided by clear human oversight~\cite{NaeemEtAl2025ThematicAnalysisArtificial}.


However, recent reflections across qualitative research have highlighted tensions between automation, ethics, and methodological integrity, stressing that AI tools should maintain interpretive agency and epistemic transparency~\cite{SchroederEtAl2025LargeLanguageModels}. A shared view has emerged in this regard covering that LLMs can complement qualitative research only within reflexive frameworks that strongly preserve human control do not replace researcher reasoning.


Despite attention to the methodological use of LLMs, empirical evidence about their role in TA remains underexplored in SE. Most prior work has evaluated technical capabilities or proposed conceptual guidelines, yet few studies have captured how researchers themselves experience and evaluate LLM assistance during the interpretive phases of TA. This study responds to that gap by examining collective reflections by experienced researchers on the perceived opportunities, limitations, and methodological tensions that arise when LLMs intersects with human interpretation in TA.



\section{Towards LLM-Assisted Thematic Analysis}
\label{sec:llmta}

\begin{figure}[ht!]
  \centering
  \includegraphics[width=8.7cm]{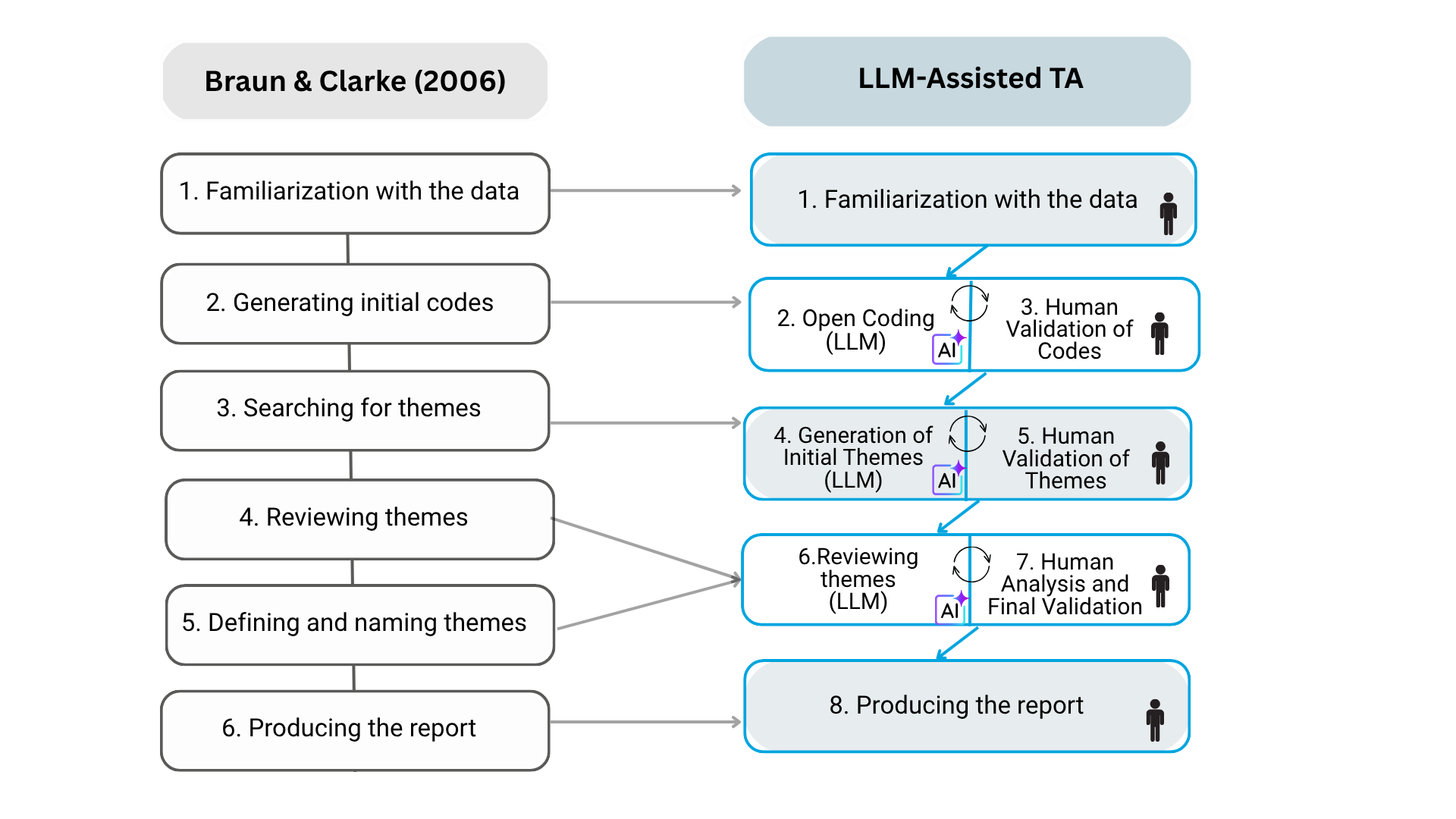}
  \caption{Overview of LLM-Assisted TA (adapted from~\cite{BraunClarke2006UsingThematicAnalysis})}
  \label{fig:llm-assisted-ta}
\end{figure}

This section conceptualizes the LLM-Assisted Thematic Analysis, which served as a structured process to guide participants’ reflections. Rather than prescribing a fixed analytical procedure, it provided a shared reference model that helped organize discussion and ensure comparability across groups as they considered how LLMs could support each stage of TA. 
Building on Braun and Clarke’s six-phase framework \cite{BraunClarke2006UsingThematicAnalysis}, this work reorganizes it into five iterative phases, consolidating overlapping steps while embedding LLMs as analytical partners through cyclical human–AI collaboration. Figure~\ref{fig:llm-assisted-ta} illustrates this alternating process between computational assistance and human interpretation, emphasizing that meaning-making remains the researcher’s responsibility.


\subsection{Phase 1: Familiarization with the Data}
This phase corresponds to Braun and Clarke’s \cite{BraunClarke2006UsingThematicAnalysis} initial step of \textit{Familiarization with the Data} and it would be conducted exclusively by human researchers. Researchers immerse themselves in the qualitative material (reading and rereading excerpts, notes, and transcripts) to develop a holistic understanding of the data. This process builds contextual awareness and theoretical sensitivity, preparing researchers to engage critically with the process.


\subsection{Phase 2: Open Coding and Human Validation}
This phase covers LLM-assisted open coding and subsequent human validation, corresponding to Steps~2 and~3 of the LLM-assisted TA illustrated in Figure~\ref{fig:llm-assisted-ta}.


In Step 2, \textit{Open Coding (LLM)},
The model performs an initial round of open coding on pre-classified qualitative excerpts, generating descriptive and data-grounded codes. The prompting strategy follows a modular design consisting of five components:
(1) research context that defines the analytical scope and theoretical constructs;
(2) data source specifying input format (e.g., interview quotes labeled by construct);
(3) coding task directing the model to conduct open coding according to Braun and Clarke’s principles;
(4) quality criteria for codes (e.g., descriptive accuracy, avoidance of premature abstraction); and
(5) output format yielding codebook (e.g., table with Quote, Code, Justification) to ensure traceability.

Following this automated stage, in Step 3, \textit{Human Validation of Codes}, researchers critically review the LLM-generated codes to assess interpretive fidelity and theoretical alignment. Validation involves verifying code–quote relations, merging redundant entries, and refining labels where needed. The objective is to preserve the authenticity of the data and sustain the researcher’s interpretive authority while benefiting from automated assistance. Building on these validated codes, the next phase focuses on identifying thematic patterns, where LLM assistance is again combined with human interpretation to generate and refine initial themes.

\subsection{Phase 3: Generation of Initial Themes and Human Validation}
This phase focuses on identifying thematic patterns from validated codes, combining LLM assistance with human interpretation. It comprises Steps 4 and 5 of the LLM-assisted TA, illustrated in Figure \ref{fig:llm-assisted-ta}, as described below.

Step 4, \textit{Generation of Initial Themes (LLM)}, is responsible for identifying higher-level patterns across the validated codes. The LLM receives these codes and groups semantically related ones into candidate themes. The prompt mirrors the logic used in open coding and includes:
(1) research context situating the task within study objectives and constructs;
(2) data source referencing validated codes from the prior phase;
(3) generation task instructing the model to cluster related codes into coherent and data-grounded themes;
(4) criteria for theme quality, including internal coherence, distinction, and interpretability; and
(5) output format producing a table with information about the themes (e.g., Theme, Description, Codes, Quotes, Participants, and optionally Relation to Construct).
The output offers a structured overview of candidate themes that accelerates hypothesis generation about data patterns.

In Step 5, \textit{Human Validation of Themes}, researchers review and refine the LLM-generated themes, ensuring conceptual coherence and alignment with research aims. They assess the clarity of thematic boundaries and rearticulate theme labels when needed. This iterative dialogue maintains theoretical depth while taking advantage of the model’s organizational capacity.

\subsection{Phase 4: Reviewing Themes}
This phase focuses on revisiting and improving the thematic structure based on insights from previous phases. It corresponds to the \textit{Defining and Naming Themes} stage of Braun and Clarke’s framework~\cite{BraunClarke2006UsingThematicAnalysis}, as shown in Figure~\ref{fig:llm-assisted-ta}. In the context of qualitative analysis assisted by LLMs, studies have reported issues such as excessive segmentation and over-categorization~\cite{ZenimotoEtAl2024CodingOpenEndedResponses}, which tend to occur in the early stages of qualitative analysis, particularly during text segmentation and initial categorization. This reinforces the need for a structured review phase to refine and consolidate themes.

In Step~6, \textit{Reviewing Themes (LLM)}, the issues to be reviewed emerge from the context of the ongoing thematic analysis and may differ across studies. In our case, the model was prompted to identify and merge semantically similar codes within each theme to address redundancy and improve conceptual coherence. The prompt includes:
(1) task description that instructs the model to identify and merge semantically similar codes within a theme;
(2) guidelines defining how to ensure semantic proximity, theoretical alignment, and preservation of meaning;
(3) expected output format requiring two tables, namely Table 1 with New Code, Grouped Codes, and Reason for Grouping, and Table 2 with Original Code, New Code, and Justification for Similarity;
(4) input section that lists the codes belonging to the same theme; and
(5) theme context that provides contextual information to guide interpretation. This prompt helps the LLM act as a semantic clustering assistant, supporting theme refinement while maintaining theoretical coherence.

Finally, in Step 7, \textit{Human Analysis and Final Validation}, researchers review all groupings and relationships between themes to confirm conceptual and analytical consistency. They assess whether the thematic map addresses the research questions, aligns with study objectives, and authentically represents participants’ voices. This stage reaffirms that interpretive authority rests with the researcher.

\subsection{Phase 5: Producing the Report}
The final phase corresponds to \textit{Reporting} phase, as illustrated in Figure~\ref{fig:llm-assisted-ta}. Researchers synthesize the validated themes into analytical narratives that address the research questions and connect empirical findings to the theoretical framework. This process prioritizes transparency, reflexivity, and faithful representation of participants’ voices. While LLMs may be used to support reporting (with proper human oversight), we considered this usage out of the scope of assisting core methodological TA phases and of our investigation.

\subsection{Concrete Example}
\label{sec:concrete_example}


To make the LLM-Assisted TA conceptualization more tangible, this section illustrates how it can be applied in practice within a real qualitative study conducted by the authors.
The aim is to exemplify, step by step, how researchers can integrate LLM assistance into different phases of TA while maintaining interpretive depth and methodological transparency.
The example draws on authentic interview excerpts from an ongoing study that explores how software engineers develop interpersonal and reflective abilities in professional contexts.


\subsubsection*{\textbf{Phase 1: Familiarization with the Data}}
As in any qualitative analysis, the process began with immersion in the data.
Before the model was ever invoked, researchers manually read and organized interview excerpts related to Observational Learning, a construct from Social Cognitive Theory (SCT) \cite{bandura2001social}.
This construct (learning by observing others’ behaviors and consequences) provided a conceptually bounded subset for demonstration.
At this stage, the goal was to understand the context deeply enough to later recognize when the model’s suggestions aligned or diverged from authentic meanings.
All excerpts were reviewed for anonymity, clarity, and relevance, laying the interpretive foundation for the next phase.

\subsubsection*{\textbf{Phase 2: Open Coding and Human Validation}}

This phase marked the first collaborative moment between LLM and human.

\paragraph{Open Coding (LLM)}
A structured prompt introduced the study’s context and asked GPT-4 to generate concise, descriptive codes for each excerpt, accompanied by brief justifications.
The result was a neatly formatted codebook (such as illustrated in Table~\ref{tab:open-coding-spreadsheet}) with each line pairing a participant’s quote with its proposed codes and explanations.

\paragraph{Human Validation of Coding}
As expected, the automation raised interpretive questions.
Researchers reviewed every code, merging redundant ones, rephrasing vague labels, and discarding those that oversimplified nuanced experiences.
This step was less about correction and more about dialogue, an opportunity to see how computational suggestions could stimulate, but not dictate, analytical thinking.
By the end of this phase, the team had a validated set of descriptive codes ready for thematic synthesis.

\begin{table*}[ht!]
\centering
\scriptsize
\caption{LLM-generated codes and justifications for excerpts during the open coding phase (Phase~2).}
\label{tab:open-coding-spreadsheet}
\renewcommand{\arraystretch}{1.25}
\setlength{\tabcolsep}{2.5pt}
\begin{tabular}{p{0.07\textwidth} p{0.26\textwidth} p{0.31\textwidth} p{0.27\textwidth}}
\toprule
\textbf{Quote ID} & \textbf{Excerpt} & \textbf{Code(s)} & \textbf{Justification} \\ 
\midrule

\textbf{S1\#G26\#P04} &
“I had an experience with a previous management [...] managed through chaos [...] there is no soft skill in such an environment.” &
Observation of toxic leadership as a negative model; \newline
Perception of the absence of soft skills in professional practice. &
The participant observes a technical leader who fails to apply soft skills, learning from the negative example. \\

\addlinespace[4pt]
\textbf{S2\#G27\#P04} &
“She had false empathy, it wasn’t genuine.” &
Recognition of false empathy; \newline
Observation of inauthentic managerial behavior. &
Learning occurs through the perception of lack of authenticity in leadership behavior. \\

\addlinespace[4pt]
\textbf{S3\#G28\#P04} &
“The leaders who are at the front are showing the way [...] the leader acts in an incorrect way.” &
Influence of leaders’ behavior as a model for new professionals; \newline
Learning from negative examples of leadership. &
Observing inadequate behaviors from leaders serves as learning about what should not be done. \\

\bottomrule
\end{tabular}
\end{table*}

\subsubsection*{\textbf{Phase 3: Generation of Initial Themes and Human Validation}}

With coding complete, attention shifted toward generation of themes.

\paragraph{Generation of Initial Themes (LLM)}
The validated codes were supplied back to the model, which was prompted to group semantically related items into candidate themes and describe each one succinctly.
Among the results were two clear thematic directions: Learning by Contrast with Negative Models and Positive Modeling and Inspiration from Successful Examples (see Table \ref{tab:initial_themes}).
The model captured distinctions that mirrored the logic of SCT, learning through both imitation and contrast.

\paragraph{Human Validation of Themes}
Researchers examined these preliminary themes, checking whether each one truly captured shared meaning or merely reflected surface similarity.
Ambiguous or unassigned codes were revisited manually.
This iterative exchange showed that while the model could accelerate theme generation, interpretive coherence still depended on the researcher’s theoretical and contextual awareness.


\begin{table*}[t]
\centering
\scriptsize
\caption{Initial themes and related codes generated and validated in Phase~3 (Generation and Validation of Initial Themes).}
\label{tab:initial_themes}
\begin{adjustbox}{max width=\textwidth}
\begin{tabularx}{\textwidth}{
  >{\RaggedRight\arraybackslash}p{0.10\textwidth} 
  >{\RaggedRight\arraybackslash}p{0.15\textwidth} 
  >{\RaggedRight\arraybackslash}p{0.31\textwidth} 
  >{\RaggedRight\arraybackslash}p{0.12\textwidth} 
  >{\RaggedRight\arraybackslash}p{0.22\textwidth} 
}
\toprule
\textbf{Theme} & \textbf{Description} & \textbf{Associated Codes} & \textbf{Quotes / Participants} & \textbf{Relationship with the Construct} \\
\midrule

\textbf{Learning by Contrast with Negative Models} &
Represents situations in which engineers learned by observing inappropriate behaviors of leaders or colleagues, identifying what should not be done. &
Observation of toxic leadership as a negative model; \newline
Perception of the absence of soft skills in professional practice; \newline
Recognition of falseness in empathetic behavior; \newline
Observation of inauthentic managerial behavior; \newline
Learning through the negative example of teamwork; \newline
Observation of negative consequences as a source of learning; \newline
Negative models reinforcing desirable behaviors; \newline
Conscious rejection of negative models. &
['S1', 'S2', 'S3', 'S12', 'S13', 'S14', 'S17'] \newline
['P04', 'P15', 'P17', 'P18'] &
Participants demonstrate attention and retention when observing dysfunctional behaviors in the professional environment, learning to avoid them. The motivation to act differently reinforces observational learning through contrast. \\

\addlinespace[4pt]
\textbf{Positive Modeling and Inspiration from Successful Examples} &
Covers experiences in which engineers observed colleagues, leaders, or public figures and adopted positive strategies in their professional conduct. &
Influence of leaders’ behavior as a model for new professionals; \newline
Direct inspiration from a colleague as a positive model; \newline
Adoption of observed interpersonal strategy; \newline
Direct observation of an inspiring leader; \newline
Internalization of advice as a model for conduct; \newline
Learning from experienced colleagues. &
['S3', 'S18', 'S19', 'S22', 'S24'] \newline
['P04', 'P18', 'P19', 'P24', 'P27'] &
This theme reflects the observation of positive models (attention), internalization of behaviors (retention), attempts at reproduction, and inspiration (motivation), which are central elements of the construct Observational Learning. \\
\bottomrule
\end{tabularx}
\end{adjustbox}
\end{table*}

\subsubsection*{\textbf{Phase 4: Reviewing Themes}}

Once initial themes were identified, the focus turned to conceptual precision.

\paragraph{Reviewing Themes (LLM)}
A new prompt instructed the model to look within each theme for overlapping or redundant codes and propose semantic groupings, each justified in plain language.
This step served as a form of assisted conceptual clean-up (helping researchers visualize which codes might belong together). The expected outputs were two tables: 
(i) consolidated codes with their original grouped codes and justification, and 
(ii) a mapping of each original code to its consolidated version, shown in
Table~\ref{tab:semantic-grouping-output}.

\paragraph{Human Analysis and Final Validation}
The research team then revisited the model’s suggestions with caution, accepting only those consolidations that preserved interpretive richness.
Some groupings clarified relationships; others, however, risked erasing meaningful distinctions.
Through this reflective filtering, the final thematic map gained coherence without losing the diversity of perspectives embedded in the data.

\begin{table}[ht!]
\centering
\scriptsize
\caption{Semantic code grouping results during Phase~4.}
\label{tab:semantic-grouping-output}
\renewcommand{\arraystretch}{1.15}
\setlength{\tabcolsep}{2pt}
\begin{tabular}{p{0.18\columnwidth} p{0.23\columnwidth} p{0.23\columnwidth} p{0.32\columnwidth}}
\toprule
\textbf{Theme} & \textbf{New Code} & \textbf{Grouped Codes} & \textbf{Reason / Participant(s)} \\
\midrule

\textbf{Learning Facilitated by Formal Structures and Events} &
\textbf{Learning Guided by Individual Models (Mentors and Leaders)} &
Learning by observation in a mentoring context; \newline
Active search for leadership models. &
Both refer to learning guided by specific individuals—mentors or leaders—through intentional observation and imitation. /
(P06, P09) \\

\addlinespace[4pt]
\textbf{Learning Facilitated by Formal Structures and Events} &
\textbf{Observational Learning in Structured Educational Settings (Trainings and Events)} &
Learning by observation in structured contexts (training); \newline
Observation of practices in corporate training sessions. &
Both describe learning occurring in formal, planned educational settings such as professional events or trainings. /
(P09, P11) \\

\bottomrule
\end{tabular}
\end{table}

\subsubsection*{\textbf{Phase 5: Producing the Report}}

The last phase returned fully to human interpretation.
Validated themes were woven into an analytical narrative and visual map that connected constructs, themes, and representative excerpts.
Reporting was treated not as an endpoint but as part of the analytic reasoning, an opportunity to reflect on how the hybrid process shaped understanding.
At this stage, the LLM’s role concluded, and meaning-making was once again exclusively human work.

\section{Research Design}
\label{sec:method}

The following subsections describe the workshop’s organization and participants (Section \ref{sec:procedure}) and the data collection and analysis procedures (Section \ref{sec:data_collection}).

\subsection{Workshop's Overview}
\label{sec:procedure}

The study was conducted through a structured reflective workshop designed to elicit collective perceptions from experienced  researchers on the integration of LLMs into TA. The session took place during the International Software Engineering Research Network (ISERN) 2025 meeting, held in September in Hawaii, USA. ISERN provides a long-standing collaborative forum for advancing empirical research methods and methodological reflection within the SE community.
The 60-minute workshop followed a collaborative format that encouraged open discussion and shared methodological reasoning. It began with a short introduction situating the rapid growth of LLM-assisted qualitative research and highlighting the need for transparent and responsible methodological practices. The session was facilitated by the last author, who presented the LLM-Assisted TA conceptualization and illustrated its phases with the concrete example explained in Section \ref{sec:concrete_example}.


After the presentation, the 25 participants were divided into five groups of roughly equal size. Each group worked collaboratively and deeply focused for 30 minutes on a large A2 canvas using color-coded post-its: opportunities (green), limitations (yellow), and recommendations (blue). The canvases were divided into three columns (open coding, generation of initial themes, and reviewing themes) to serve as a structure for participants to reflect systematically on methodological, ethical, and practical aspects of LLM integration into the analytical phases of TA. 


The workshop concluded with a plenary in which each group summarized its key points about how using LLMs for TA could be improved. In particular, the facilitator synthesized the shared insights, encouraging participants to connect their reflections to broader issues of rigor, reproducibility, and interpretive integrity in LLM-assisted qualitative research. The facilitator adopted a neutral and supportive stance throughout, fostering open exchange rather than guiding participants toward consensus.


\subsection{Data Collection and Analysis}
\label{sec:data_collection}

\begin{figure}[ht!]
  \centering
  \includegraphics[width=0.8\linewidth]{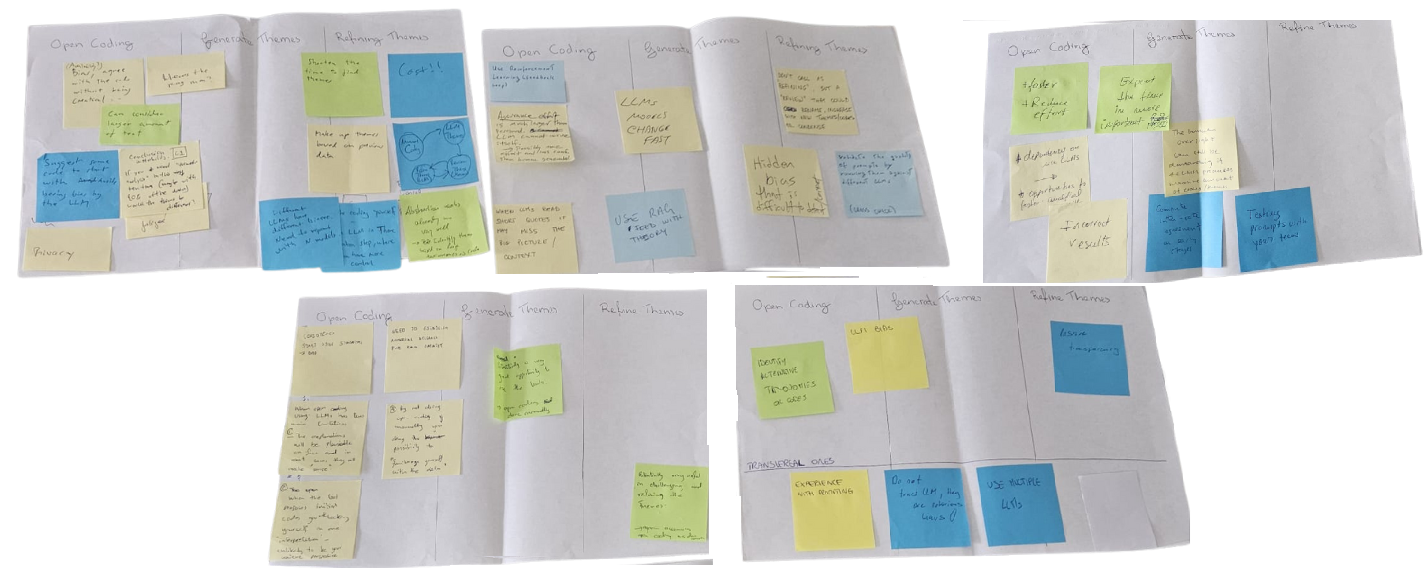}

  \caption{Canvases built in the workshop}
  \label{fig:photo}
\end{figure}

The collected material for this study consisted of the completed A2 canvases produced by each group. Each canvas contained color-coded post-its representing reflections on opportunities, limitations, and recommendations for the three key analytical phases of the LLM-Assisted TA. After the session, all canvases were photographed (see Figure \ref{fig:photo}), and the content was transcribed into a spreadsheet for further analysis.

The transcribed material was examined through a structured descriptive synthesis aimed at identifying collective patterns in participants’ reflections. Each post-it note was treated as a unit of analysis and associated with its respective analytical phase (open coding, theme generation, theme reviewing), or labeled as cross-phase when researchers judged it relevant to all phases, and with its color category (opportunities, limitations, recommendations). The spreadsheet was organized along these two dimensions to enable systematic comparison across groups and phases. The analysis unfolded iteratively. First, all reflections were reviewed to ensure transcription accuracy and contextual consistency. Next, similar or complementary ideas were grouped to identify recurring topics and cross-cutting tensions. Finally, these groupings were synthesized into higher-level categories that summarized how researchers conceptualized the potential, risks, and boundaries of human–LLM collaboration in TA. The procedure was interpretive and exploratory rather than evaluative, seeking to distill the essence of participants’ perceptions rather than quantify consensus or assess model performance. Analytical transparency was maintained by documenting each step of transcription, categorization, and synthesis, and by preserving traceable links between the raw material and the synthesized outputs. All collected materials, including the consolidated spreadsheet, are available in our open repository~\cite{zenodo}.

\section{Results}
\label{sec:results}

\subsection{Open Coding}

\subsubsection{Opportunities}

Participants perceived some advantages in using LLMs during the open coding phase, particularly related to efficiency, scalability, and creative exploration. Some noted that these tools could \textit{foster and reduce effort}, highlighting the potential of automation to alleviate repetitive tasks and accelerate the identification of patterns across large datasets. Others emphasized that LLMs could \textit{consider larger amounts of text}, allowing researchers to process data at a scale that would be impractical through manual coding alone. This was associated with expectations for greater analytical breadth, with one group describing the possibility to \textit{expect the future in more important ways}, suggesting that LLMs might help uncover latent or emergent dimensions of qualitative data.

Finally, participants mentioned that the models could \textit{identify alternative taxonomies or codes}, pointing to their capacity to propose diverse conceptual categorizations. This issue indicates that participants recognized the potential of LLMs to inspire new interpretive directions and support theory-building in qualitative analysis.

\subsubsection{Risks and Limitations}

Across several groups, participants raised concerns about growing dependency on LLMs, whether technical, cognitive, or interpretive, and how this could reduce researchers’ opportunities to exercise critical evaluation and maintain direct engagement with data. Comments such as \textit{increase dependency on LLMs} and \textit{decrease opportunities to foster evaluating skills} captured this apprehension, emphasizing that excessive automation might weaken human analytical judgment and reflexivity.

A reinforced orientation emerged around the importance of maintaining manual engagement in the open coding phase. Two groups explicitly argued that skipping manual coding undermines the researcher’s familiarity with the data: \textit{by not doing open coding manually you are denying the possibility of familiarising yourself with the data.} Participants viewed this phase as fundamental for developing theoretical sensitivity and contextual understanding, cautioning that over-automation could weaken the interpretive foundation upon which subsequent analytical stages rely.

Participants also expressed doubts about the reliability and stability of LLM-generated codes. Some mentioned \textit{incorrect results} and \textit{conclusion instability}, questioning whether similar datasets would lead to consistent thematic outcomes. Others noted that \textit{LLMs cannot assure themselves}, implying that verifying outputs may require \textit{more effort and less confidence than human-generated results}. One group added that \textit{the human oversight can still be demanding}, as models often produce a massive number of codes, increasing the workload required for review and validation. Contextual interpretation emerged as another major limitation. One group observed that \textit{when LLMs read short quotes, they may miss the big picture or context}, producing explanations that appear plausible but lack depth. This view was echoed in a collective note warning that \textit{by not doing open coding manually you are denying the possibility of familiarising with the data.}

Concerns about analytical bias and authenticity were also present, with statements such as \textit{agree with the code without being analytical} and \textit{where’s the real human?} reflecting participants’ discomfort with delegating interpretive authority to machines. Additional risks mentioned included \textit{creativity fatigue}, \textit{privacy issues}, and the need to \textit{establish numerical accuracy for each dataset.} These reflections show that while LLMs can support and accelerate open coding, they simultaneously introduce methodological and epistemic trade-offs. Maintaining human interpretive agency and contextual sensitivity was seen as essential to preserving rigor and authenticity in qualitative research.

\subsubsection{Suggestions and Safeguards}

Participants proposed a few practical suggestions for integrating LLMs responsibly into the open coding process. One recurring idea was the importance of maintaining a comparative dynamic between human and AI-generated outputs. A group recommended comparing insights produced by LLMs with those generated by human researchers as a form of ongoing triangulation, ensuring that automation complements rather than replaces human interpretation. Another suggestion involved the use of reinforcement and iterative feedback loops, in which researchers review model outputs and provide corrective input. This cyclical interaction was seen as a way to improve contextual alignment and preserve human reasoning at the center of the analysis. One group also suggested guiding the model through initial examples before running the full analysis, as a way to orient the LLM toward the research context and reduce misalignment.

These contributions point toward a reflective and dialogical use of LLMs in open coding, where human analysts retain interpretive control and use AI assistance to broaden analytical insight rather than automate interpretation.

\subsection{Generation of Initial Themes}

\subsubsection{Opportunities}

Participants identified opportunities for using LLMs during the generation of initial themes, particularly emphasizing efficiency, pattern recognition, and support for analytical abstraction. Some noted that LLMs could \textit{shorten the time to find themes}, highlighting the potential of automation to accelerate the synthesis of codes into higher-level conceptual patterns.

Some participants observed that \textit{LLMs can identify repeating patterns easily and give codes and save complexity}, suggesting that models can effectively assist in detecting recurring relationships among codes that may not be immediately visible to researchers. Others mentioned that \textit{abstraction works already very well}, indicating that LLMs can facilitate the conceptual step of grouping descriptive codes into coherent themes, especially when working with large taxonomies or extensive datasets. One group also considered LLMs \textit{potentially very useful in challenging and validating the themes}, pointing to their role not only in supporting automation but also in promoting reflection and comparison of different interpretations. Similarly, comments such as \textit{improve discussing open coding into patterns} and \textit{expect the future in more important ways} conveyed the idea that LLMs can stimulate broader analytical thinking and help researchers envision alternative thematic structures.

Overall, participants perceived that, when used with human oversight, LLMs can enhance the process of moving from fragmented codes to structured themes, improving analytical coherence, speed, and the capacity for reflective synthesis.

\subsubsection{Risks and Limitations}

Only a few risks were identified for this phase, reflecting a generally more optimistic view compared to open coding. Some participants cautioned that \textit{the human oversight can still be demanding}, especially because LLMs may \textit{produce massive amounts of codes or themes} that require extensive review. This was interpreted as a reminder that efficiency gains in automation do not eliminate the need for rigorous human validation. Concerns about bias also appeared, with participants mentioning \textit{LLM bias} and the risk of generating \textit{themes based on previous data}, suggesting that models may reproduce or reinforce patterns learned from prior training data rather than staying grounded in the current dataset. Additionally, one group noted practical limitations such as \textit{cost}, emphasizing that large-scale LLM use may introduce financial and computational constraints. Hence, participants acknowledged that while LLMs can streamline the generation of themes, researchers must remain attentive to potential distortions, overproduction of content, and the continuing necessity of careful human oversight.

\subsubsection{Suggestions and Safeguards}

A single group proposed a few targeted safeguards for using LLMs in the generation of initial themes. They emphasized that \textit{different LLMs have different biases}, suggesting that repeating the process with more than one model could improve reliability. The same group also recommended \textit{doing the coding manually and involving the LLM only in the theme generation step}, where human control over abstraction can be more effectively maintained. As a practical illustration, they outlined a simple workflow: \textit{Manual Coding → LLM Themes → Refine Themes → Review Themes (LLM)}. 

\subsection{Reviewing Themes}

In the Phase 4, which corresponds to the reviewing stage of the LLM-Assisted TA, participants’ reflections suggested that human–AI collaboration can support the refinement of themes while preserving interpretive depth. Some participants saw the potential to \textit{shorten the time to review themes}, viewing automation as a means to enhance efficiency without replacing human oversight. Others argued that this phase should not be called \textit{refining} but rather \textit{reviewing}, since it can promote \textit{new empathy or consensus} among researchers. Finally, one group stressed the need to \textit{ensure transparency} in documenting revisions and decisions, reinforcing that interpretive authority must remain with the human analyst.

\subsection{Cross-Phase Reflections}

Participants shared a set of reflections considered applicable across all Phases of the LLM-Assisted TA. The discussion addressed general methodological and ethical aspects of integrating LLMs rather than being limited to a specific analytical stage. For example, one group suggested \textit{testing prompts with research teams} and \textit{validating their quality by using different LLMs}, proposing cross-model comparison as a way to enhance reliability and reduce bias. The same group emphasized the value of \textit{using multiple LLMs} or incorporating \textit{RAG feeds with theory} to improve theoretical alignment and maintain analytical traceability throughout the process.

Participants also identified transversal risks that cut across all stages. These included the \textit{rapid evolution of LLM models}, which may compromise reproducibility, and the presence of \textit{hidden biases that are difficult to detect or correct}, posing challenges for transparency and ethical use. Some noted that effective prompting demands specific expertise, pointing to a learning curve in researchers’ \textit{experience with prompting}. Finally, a concise yet striking warning — \textit{do not trust LLMs, they are notorious liars} — summarized the shared understanding that human oversight remains essential for ensuring credibility and interpretive rigor when working with LLMs.
\section{Discussion}
\label{sec:discussion}
The reflections brought by this study show how experienced researchers in SE are beginning to conceptualize the methodological consequences of integrating LLMs into TA. Participants expressed cautious optimism since they valued efficiency but emphasized that interpretive depth, reflexivity, and contextual awareness must remain central to analytical rigor. Across all phases of the LLM-Assisted TA, the consensus was that the value of automation depends on how humans design, validate, and interpret model outputs.

Initially, participants acknowledged that LLMs can ease repetitive work and accelerate coding, confirming prior observations that automation can expand analytical reach~\cite{BanoEtAl2023LargeLanguageModels, BarrosEtAl2025LargeLanguageModel}. Yet, they also viewed immersion in the data as indispensable for developing theoretical sensitivity~\cite{BraunClarke2006UsingThematicAnalysis}. Rather than replacing manual coding, LLMs were seen as useful for exploration and triangulation (i.e., tools to compare perspectives, not to substitute human reasoning). These reflections reinforce the importance of hybrid workflows where automation complements rather than replaces interpretation.

We noticed concerns about instability, bias, and opacity that highlight  methodological rigor in LLM-assisted analysis shifts from code accuracy to the transparency of human–LLMs interaction. Echoing De Paoli~\cite{Paoli2024PerformingInductiveThematic} and Baltes et al.~\cite{BaltesEtAl2025GuidelinesEmpiricalStudies}, researchers stressed documenting prompts, model versions, and validation steps as essential to analytical traceability. Reliability, in this view, is not a property of the model but of the documented interaction between researcher and model, an idea increasingly important to open and reproducible empirical research \cite{mendez2020open}.

We also found that researchers admit LLMs as analytical partners capable of suggesting new framings or patterns but requiring informed human mediation. Hybrid approaches (alternating between model suggestions and researcher review) were viewed as most effective. This perspective aligns with Trinkenreich et al.~\cite{TrinkenreichEtAl2025GetTrainBe}, who argue that responsible engagement with AI tools requires maintaining interpretive control. The emerging skill set includes prompt literacy and validation awareness, pointing to new methodological competencies for empirical SE.

In summary, the findings point out that LLMs make existing challenges (such as reflexivity, traceability, and ethical accountability) more visible rather than replacing them. Responsible integration will require explicit reporting of prompts, model configurations, and researcher interventions, as well as shared community resources. Therefore, LLMs may not redefine qualitative research, but they intensify its methodological demands, inviting SE scholars to rethink how rigor and judgment are enacted in human–LLMs analysis.

\section{Final Remarks}
\label{sec:conclusion}

Large Language Models (LLMs) are rapidly influencing how empirical researchers in Software Engineering (SE) approach qualitative research. This study contributes an empirical perspective on how experienced researchers perceive the methodological opportunities and challenges of integrating LLMs into Thematic Analysis (TA). Their reflections portray a community cautiously engaging with automation, acknowledging its potential to improve efficiency and analytical reach, while reaffirming the irreplaceable role of human interpretation, contextual awareness, and reflexivity.

The study was conducted during the International Software Engineering Research Network (ISERN) 2025 meeting, through a reflective workshop structured around the LLM-Assisted TA, an adaptation of Braun and Clarke’s proposal \cite{BraunClarke2006UsingThematicAnalysis} that conceptually embeds human–AI collaboration across analytical phases. This conceptualization provided an preliminary reference for participants to reason systematically about how LLMs might support or complicate open coding, theme generation, and theme reviewing. The discussions indicated that rigor in hybrid human–AI analysis increasingly depends on transparent documentation of prompts, validation procedures, and researcher interventions. They also highlighted emerging methodological competencies (such as prompt literacy, interpretive validation, and ethical awareness) as necessary for responsible LLM use in empirical SE.

As with any interpretive study, this work is bounded by its design and has limitations. The reflections were collected from a single collaborative workshop involving a focused group of senior empirical SE researchers. The goal was not statistical generalization but the elicitation of collective reasoning about methodological practice. Consequently, the findings capture an informed yet situated view of how experienced researchers currently conceptualize the role of LLMs in qualitative analysis. Future studies could broaden this understanding through comparative or longitudinal designs, empirical evaluation of LLM-assisted outputs, and replication in diverse research communities. Overall, the reflections emerging from this ISERN work suggest that LLMs do not replace qualitative methodology but instead prompt the SE community to revisit core principles of rigor, transparency, and interpretive accountability in a human–LLM research scope.


\section*{Acknowledgments}
 We express our gratitude to CNPq (Grant 312275/2023-4), FAPERJ (Grant E-26/204.256/2024), Kunumi Institute, and Stone Co. for their generous support. We also thank the ISERN for providing a distinguished forum to steer advancements in empirical methods.

\bibliographystyle{ACM-Reference-Format}
\bibliography{references}

\end{document}